%% file: LLgosam2.tex
\documentclass{PoS}

\usepackage{cite}

\newcommand{\gosam}{{\textsc{GoSam}}}
\newcommand{\gosamv}{{\textsc{GoSam-2.0}}}
\newcommand{\GOSAM}{{\textsc{GoSam}}}
\newcommand{\QGRAF}{{\texttt{QGRAF}}}
\newcommand{\FORM}{{\texttt{FORM}}}
\newcommand{\SPINNEY}{{\texttt{spin\-ney}}}
\newcommand{\HAGGIES}{{\texttt{hag\-gies}}}
\newcommand{\SAMURAI}{{\textsc{Sa\-mu\-rai}}}
\newcommand{\GOLEMVC}{{\textsc{Go\-lem95C}}}
\newcommand{\NINJA}{{\textsc{Ninja}}}

\newcommand{\bea}{\begin{eqnarray*}}
\newcommand{\eea}{\end{eqnarray*}\noindent}
\newcommand{\be}{\begin{equation}}
\newcommand{\ee}{\end{equation}\noindent}

\def\S{\mathcal{S}}

\newcommand{\mrm}[1]{\mathrm{#1}}
\newcommand{\bite}{\begin{itemize}}
\newcommand{\eite}{\end{itemize}}

\title{Automated one-loop calculations with GoSam 2.0}

\ShortTitle{Automated one-loop calculations with GoSam 2.0}

\author{H.~van Deurzen, N.~Greiner, \speaker{G.~Heinrich}, G.~Luisoni,
E.~Mirabella,
T.~Peraro,  J.~Schlenk, J.~F.~von~Soden-Fraunhofen\\ 
Max Planck Institute for Physics, Munich\\
\email{\{hdeurzen,
 greiner,
 gudrun,
 luisonig,
 mirabell,
 peraro,
 jschlenk,
 jfsoden\}@mpp.mpg.de}
 }

\author{P.~Mastrolia\\ Max Planck Institute for Physics, Munich, 
and Dipartimento di Fisica e Astronomia, Universit\`a di Padova, and INFN Sezione di Padova, 
Padova, Italy\\
\email{Pierpaolo.Mastrolia@cern.ch}}

\author{G.~Ossola\\ New York City College of Technology, City University of New York, USA\\
\email{gossola@citytech.cuny.edu}}

\author{F.~Tramontano\\  Dipartimento di Scienze Fisiche, Universit\`a degli studi di Napoli and INFN, Sezione di Napoli, 
Napoli, Italy\\
\email{Francesco.Tramontano@cern.ch}}

\abstract{We present the version 2.0 of the program {\sc GoSam}, which  
is a public program package to compute one-loop corrections 
to multi-particle processes.
The extended version of the ``Binoth-Les-Houches-Accord'' interface 
to Monte Carlo programs  is also implemented. 
This allows a large flexibility regarding the combination of the code with 
various Monte Carlo programs to produce fully differential NLO results, 
including the possibility of parton showering and hadronisation.
We describe the new features of the code and illustrate the wide range of applicability 
for multi-particle processes at NLO, 
both within and beyond the Standard Model.}

\FullConference{ Loops and Legs in Quantum Field Theory - LL 2014,\\
		 27 April - 2 May 2014 \\
		 Weimar, Germany }

\begin{document}

\section{Introduction}

\label{sec:intro}
\input intro

\section{Workflow of the program}
\label{sec:overview}
\input overview

\section{New features}
\label{sec:newfeatures}
\input{newfeatures}

\section{Installation and usage}
\label{sec:instantuse}

\subsection{Installation}
\label{sec:install}
\input installation

\subsection{Using \gosam}
\label{sec:usage}
\input usage

\section{Applications}
\label{sec:examples}
\input examples

\section{Conclusions}
\GOSAM-2.0 is a flexible tool to calculate one-loop QCD  corrections 
both within and beyond
the Standard Model, as well as  electroweak corrections. 
The new version of the program offers a wide range of applicability. 
In particular, 
integrals where the rank exceeds the number of propagators (needed
e.g. in effective theories) are fully supported, and propagators for spin-2 particles are
implemented.  The complex mass scheme is supported, including the complexification
of the couplings, and several electroweak schemes can be chosen. 
\GOSAM-2.0 also contains a new integrand
reduction method, the integrand decomposition via Laurent expansion,
implemented in the library \NINJA{}, which leads to a considerable
gain in stability and speed.
In addition, \GOSAM-2.0 can provide 
spin-and colour correlated tree-level matrix elements.
Monte Carlo programs can be interfaced easily using the Binoth-Les-Houches-Accord, where  
the new standards are also supported.
The automated interface to  Monte Carlo programs offers the 
possibility to produce parton showered events and to compare different 
Monte Carlo event generators at the NLO level.

\section*{Acknowledgements}
\noindent We would like to thank Gavin Cullen for very fruitful collaboration 
in earlier phases of the GoSam project.
We are also grateful to Joscha~Reichel for contributions to the 
BSM application  of GoSam involving spin-2 particles.
P.M., H.v.D., G.L. and T.P. are supported by the Alexander von
Humboldt Foundation, in the framework of the Sofja Kovaleskaja Award Project
2010.
The work of G.O. was supported in part by the National Science Foundation
under Grant PHY-1068550.

\bibliographystyle{spphys}

\input LLgosam2.bbl


\end{document}

%% file: intro.tex
The development of automated NLO tools has seen much progress in
the past years, leading to 
an almost industrial production of NLO QCD predictions for multi-particle processes. 
In fact, the pre-2013  ``Les Houches wishlist'' of Standard Model processes which are 
desired to be available at NLO is meanwhile fully ticked in what concerns fixed order QCD predictions, 
moving now towards parton shower matching, electroweak corrections, interferences, NNLO corrections~\cite{Butterworth:2014efa}.

The standard at the LHC should be NLO corrections matched to a parton shower, 
ideally including the matching of different 
jet multiplicities. Flexible, automated NLO codes are a pre-requisite to achieve this goal. 
Here we will present the version 2.0 of the code \gosam{} which 
is able to produce multi-particle one-loop predictions for user-defined processes, 
both within the Standard Model and beyond.

\GOSAM{}~\cite{Cullen:2011ac,Cullen:2014yla} is a code which was designed 
to maximally exploit both the integrand
reduction for dimensionally regulated one-loop
amplitudes \cite{Ossola:2006us,Giele:2008ve}, as implemented
in \SAMURAI{} \cite{Mastrolia:2010nb}, as well as 
improved tensor reduction methods as developed in \cite{Binoth:2005ff}
and implemented in the code \GOLEMVC{}\cite{Binoth:2008uq,Cullen:2011kv,Guillet:2013msa}.
An important new feature in version 2.0 is the fact that \GOSAM{} can also be
interfaced with a new library, called \NINJA{}
\cite{Peraro:2014cba,vanDeurzen:2013saa}, 
implementing an ameliorated integrand-reduction method, where the
decomposition in terms of master integrals is achieved by Laurent
expansion through semi-analytic polynomial
divisions \cite{Mastrolia:2012bu}.  

The new version of the
program \GOSAM~\cite{Cullen:2014yla} has been used already to
produce a number of NLO predictions both
within~\cite{Greiner:2012im,vanDeurzen:2013rv,Gehrmann:2013aga,Luisoni:2013cuh,Hoeche:2013mua,Cullen:2013saa,vanDeurzen:2013xla,Gehrmann:2013bga,Dolan:2013rja,Heinrich:2013qaa}
and beyond~\cite{Cullen:2012eh,Greiner:2013gca} the Standard Model.
It contains important improvements in speed, numerical
robustness,  range of applicability and user-friendliness.
\GOSAM{} can be linked to different Monte Carlo programs via the Binoth-Les-Houches-Accord BLHA~\cite{Binoth:2010xt},
where the extended version BLHA2~\cite{Alioli:2013nda} is also implemented. 
The program can be downloaded from~{\tt http://gosam.hepforge.org}.

%% file: overview.tex
\begin{center}
\begin{figure}[htb]
\begin{picture}(100,180)
\put(90,-282){
\includegraphics[width=13.cm]{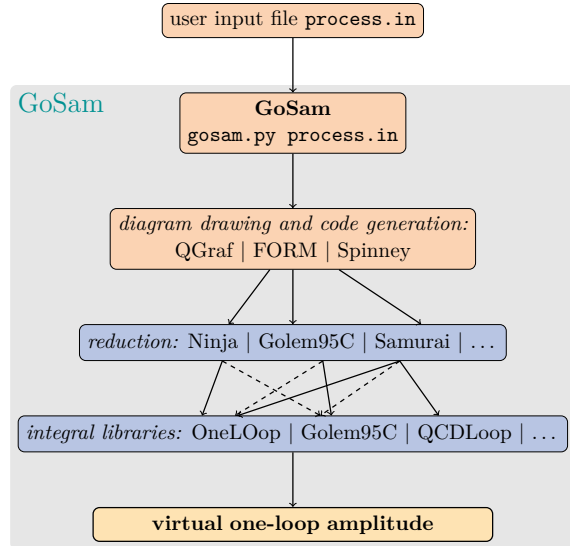}
}
\end{picture}
\caption{Basic flowchart of \GOSAM{}.}
\label{fig:flowchart}
\end{figure}
\end{center}

\GOSAM{} can be used either as a standalone code producing one-loop 
(and colour/spin correlated tree level) amplitudes, or it can be used 
in combination with a modern Monte Carlo  event generator program. 
The main workflow of \GOSAM{} is shown in Fig.~\ref{fig:flowchart}. 
In the standalone version, the user will fill out a process run
card {\tt process.in} to define the process to be calculated, 
where also further options are available.
Then the code for the virtual amplitudes
is generated by invoking {\tt gosam.py process.in}\,.\\ After running
the above command with an appropriate run card, all the files which
are relevant for code generation will be created.  The command {\tt
make source} will invoke \QGRAF~\cite{Nogueira:1991ex}
and \FORM~\cite{Vermaseren:2000nd,Kuipers:2012rf} to generate the
diagrams and algebaric expressions for the amplitudes, using also 
\SPINNEY~\cite{Cullen:2010jv} for the spinor algebra
and \HAGGIES~\cite{Reiter:2009ts} for code generation.  In
version 2.0 of \GOSAM{}, the production of optimized code 
largely relies on the new features of \FORM~version $\geq 4$.  The
command {\tt make compile} will finally compile the produced  code.

We should point out  that all the  reduction- and integral libraries 
used in \gosamv{} are included
in the program package, and the installation script  will take care of compilation and linking,
such that the user does not have to worry about dependencies.

In the mode where \GOSAM{} is used in combination with a Monte Carlo program 
providing the real radiation and infrared subtraction parts
and performing the phase space integration, 
the whole amplitude generation process can be invoked
automatically and steered by the Monte Carlo program setup, 
if the MC has the appropriate functionality.
This is most conveniently done 
via the Binoth-Les-Houches-Accord BLHA
interface~\cite{Binoth:2010xt,Alioli:2013nda}, 
which allows to link one-loop programs to Monte Carlo programs in a 
standardized way.   
This setup also allows to produce NLO results matched to a parton shower
if provided  by the Monte Carlo program.



%% file: newfeatures.tex

\subsection{Improvements in code generation}


Using largely the features provided by \FORM{} version
$\geq 4.0$~\cite{Kuipers:2012rf} to produce optimized expressions leads to more
compact code and a speed-up in amplitude evaluation of about a factor
of ten.  


In addition, the new release contains an option called {\tt diagsum} which combines diagrams
differing only by a subdiagram into one ``meta-diagram'' to be
processed as an entity. This  further reduces the number of
calls to the reduction program and therefore  increases the
computational speed. 
The {\tt diagsum}  mode is activated by default.

Another new feature in \gosamv{} is the use of numerical polarisation vectors for massless gauge bosons.  
This means that the
various helicity configurations for the massless bosons will be
evaluated numerically, based on a unique code containing generic polarisation vectors, 
rather than producing separate code for each helicity configuration.

\subsection{Improvements in the reduction}
\label{sec:ninja}

In \GOSAM{}-2.0, the amplitude decomposition can be 
obtained by a new integrand-reduction method~\cite{Mastrolia:2012bu}, 
implemented in the C++ code
\NINJA~\cite{Peraro:2014cba,vanDeurzen:2013saa}.
This novel $D$-dimensional unitarity-based algorithm is lighter than
the original integrand reduction method, because less coefficients
need to be determined, and it was found to be faster and numerically
more accurate.
The integrand-reduction library \SAMURAI~\cite{Mastrolia:2010nb,vanDeurzen:2013pja}, 
as well as the tensor reduction- and integral library
\GOLEMVC~\cite{Binoth:2008uq,Cullen:2011kv,Guillet:2013msa}
are also available in \GOSAM{}-2.0, and it is possible to switch between the different 
reduction methods ``on the fly''. 
We would like to point out that the reduction library \GOLEMVC{}
also contains an interface to integrand-reduction methods 
via tensorial reconstruction~\cite{Heinrich:2010ax}, 
where the coefficients of tensor integrals rather
than scalar integrals are reconstructed, and the tensor integrals are called from 
\GOLEMVC{}.
To rescue phase space points which do not pass the numerical precision test, 
we use \GOLEMVC{} as the default. 



\subsection{Electroweak scheme choice}
\label{sec:ewchoose}
Renormalisation within the Standard Model can be performed 
using various schemes, which also may differ in the set of  electroweak parameters 
considered as input parameters, while other electroweak parameters are derived ones.
Within \GOSAM{}-2.0, different schemes can be chosen  by setting appropriately the 
flag {\tt model.options}.
If the flag is not set in the input card, \GOSAM{}
generates a code which uses $\mrm{M_W}$, $\mrm{M_Z}$ and
$\mrm{\alpha}$ as input parameters.

\subsection{Stability tests and rescue system}
\label{sec:rescue}

For the precision analysis contained in \gosamv, and to set the trigger for the rescue system, 
we employ a hybrid method that combines the benefits in computational speed of 
testing the accuracy of the pole coefficients (called  {\it pole test}) 
with the higher reliability of testing the finite part, where the latter is achieved by
exploiting the invariance of the scattering amplitudes under an azimuthal rotation 
about the beam axis (called {\it rotation test}).  
In more detail, after computing the matrix elements, 
\gosamv{} checks the precision  $ \delta_{pole}$ 
of the single pole with the {\it pole test}. Comparing the coefficient of the single pole 
$\S_{IR}$ that can be obtained from the general structure of infrared singularities 
to the one provided by  \gosamv, which we label $\S$, we define  
\be \label{eq:exd}
\delta_{pole} = \left | ( \S_{IR} - \S{} )/ \S_{IR} \right |\, .
\ee
The corresponding estimate of the number of correct digits in the result is provided by  
$P_{pole}= - \log_{10} (\delta_{pole})$. 
The value of $ P_{pole}$ is then compared with two thresholds $ P_{high}$ and $ P_{low}$. 
If $P_{pole} >  P_{high}$, the point is automatically accepted. 
If $P_{pole} <  P_{low}$, the point is automatically discarded, 
or sent to the rescue system. If already the pole coefficient has a low precision, 
we can expect the finite part to be of the same level or worse.
In the intermediate region where $ P_{high} > P_{pole} >  P_{low}$, 
it is more difficult to determine the quality of the result solely based on the 
pole coefficients. 
Only in this case the point is recalculated using  a phase space point 
which is rotated around the beam axis to assess the numerical precision
of the original point, i.e. the 
{\it rotation test} is performed.



\subsection{New range of applicability}

\subsubsection*{Higher rank integrals}

The libraries \NINJA, \GOLEMVC{} and \SAMURAI{} all support integrals
with tensor ranks exceeding the number of propagators. 
These extensions are described in detail in
Refs.~\cite{vanDeurzen:2013pja,Guillet:2013msa,Mastrolia:2012bu} and
are contained in the distribution of \GOSAM-2.0.  The additional
integrals will be called automatically by \GOSAM{} if they occur in an
amplitude, such that the user can calculate amplitudes involving
higher rank integrals without additional effort.

\subsubsection*{Production of colour-/spin correlated tree amplitudes}

\GOSAM{} can also generate  tree level amplitudes in a spin- and colour-correlated form.
These matrix elements are particularly useful in combination with
Monte Carlo programs which use these trees to build the dipole
subtraction terms for the infrared divergent real radiation part. With
these modified tree level matrix elements \GOSAM{} is able to generate
all necessary building blocks for a complete NLO calculation. Such a
setup has been used successfully in combination with the framework of
{\sc Herwig++/Matchbox}~\cite{Butterworth:2014efa,Bellm:2013lba}.

\subsubsection*{Support of complex masses}

The integral libraries contained in the \GOSAM{} package as well as the \GOSAM{} 
code itself fully support complex masses. 
The latter are needed  for the treatment of 
unstable particles  via the   introduction of the corresponding decay width. 
A consistent treatment of complex
$W$- and $Z$-boson masses is given by the so-called complex mass scheme~\cite{Denner:2005fg}.
According to this scheme the boson masses and the weak mixing angle become
\begin{equation}
 m_{V}^2 \to \mu_{V}^2 = m_{V}^2 -i m_{V} \Gamma_{V}\; (V=W,Z)\;,
 \quad \cos^2\theta_W = \frac{\mu_W^2}{\mu_Z^2}\;.
\end{equation}
 To use the complex mass scheme, the corresponding model files \texttt{sm\_complex}
 or \texttt{smdiag\_complex} should be called, where  the latter 
 contains a diagonal unity CKM matrix.


%% file: installation.tex
The user can download the code either as a tar-ball
    or from a subversion repository at
\begin{center}
    {\tt http://gosam.hepforge.org}
\end{center}

\noindent The installation of \gosamv{} is very simple when using the installation script.
The latter can be downloaded by
\begin{flushleft}
 {\tt wget } {\tt http://gosam.hepforge.org/gosam-installer/gosam\_installer.py}
\end{flushleft}
To run the script the user should execute the following commands
\begin{flushleft}
{\tt chmod +x gosam\_installer.py}\\
{\tt ./gosam\_installer.py  [--prefix=...]}
\end{flushleft}
Upon installation, the installer will ask some questions, which are
described in more detail in the manual~\cite{gosamhome}. 
To use the default installation all the questions can be ``answered'' by pressing the {\tt ENTER} key.
The program \GOSAM{} is designed to run in any modern Unix-like environment (Linux, Mac).

%% file: usage.tex

Here we only describe the usage of \GOSAM{} in the standalone
version.  For the usage in combination with a Monte Carlo program, based
on the BLHA interface, we refer to the manual~\cite{gosamhome}.
In order to generate the matrix element for a given process the user
should create a process specific setup file, \texttt{process.in},
which we call {\em process card}.  A simple example process card for the
QCD corrections to the process $e^+e^-\to t\bar{t}$, 
where all options take default values, 
looks as follows:
\begin{center}
{\tt
\begin{tabular}{l}
in = e+,\,e-\\
out = t,\,t$\sim$\\
order = gs,0,2
\end{tabular}
}
\end{center} 
It is recommended to generate and modify a template file for the
process card. This can be done by invoking the shell command 
{\tt gosam.py -\,-template process.in}\,,\\
which generates the file \texttt{process.in} with some documentation for
all defined options. The options are filled with the default values,
and some paths are set by the installation script. 
In order to generate the Fortran code for the process specified in the
input card the user simly needs to invoke
{\tt gosam.py process.in}\,.

%% file: examples.tex
Some examples of processes where the new version of \GOSAM{} has already been used, 
in combination with various Monte Carlo programs,  to produce 
phenomenological results are shown in Table~\ref{tab:interface}. 
More details about the Higgs plus jets processes can also be found in the contribution of 
F.~Tramontano~\cite{francesco}.

\begin{table}
\caption{NLO calculations done by interfacing \GOSAM{} with different Monte Carlo programs.   
\label{tab:interface}}
\begin{center}
\begin{tabular}{|l|}
\hline
\hline
\GOSAM{} + MadDipole/MadGraph4/MadEvent\,\cite{Frederix:2008hu,Frederix:2010cj,Gehrmann:2010ry,Maltoni:2002qb,Alwall:2007st}\\
\hline
$pp\to W^+W^-$ + 2\,jets\,\cite{Greiner:2012im}\\
$pp\to \gamma\gamma$ + 1,2\,jets \,\cite{Gehrmann:2013aga,Gehrmann:2013bga} \\
SUSY QCD corrections to $pp\to \tilde{\chi}_1^0 \tilde{\chi}_1^0$ + 1\,jet \,\cite{Cullen:2012eh} \\
QCD corrections to $pp\to $ graviton($\to \gamma\gamma$) + 1\,jet \,\cite{Greiner:2013gca} \\
 \hline
 \hline
\GOSAM{} + {\sc Sherpa}\,\cite{Gleisberg:2007md,Gleisberg:2008ta}\\
\hline
$pp\to W^+W^-\,b\bar{b}$\,\cite{Heinrich:2013qaa}\\
$pp\to t\bar{t}$ + 0,1 jet (merged+shower)\,\cite{Hoeche:2013mua}\\
$pp\to H$ + 2\,jets (gluon fusion)\,\cite{vanDeurzen:2013rv}\\
$pp\to t\bar{t}H$ + 0,1 jet\,\cite{vanDeurzen:2013xla} \\
$pp\to W^+W^+ $ + 2\,jets\,\cite{gosamproc}\\
$pp\to W^\pm b\bar{b}\; (m_b\not=0)$ \,\cite{gosamproc}\\
 \hline
 \hline
\GOSAM{} + MadDipole/MadGraph4/MadEvent+{\sc Sherpa}\\
\hline
$pp\to H$ + 3\,jets (gluon fusion) \, \cite{Cullen:2013saa}\\
\hline
 \hline
\GOSAM{} + {\sc Powheg}\,\cite{Frixione:2007vw,Alioli:2010xd}\\
\hline
$pp\to HW/HZ$ + 0,1 jet (merged+shower)\,\cite{Luisoni:2013cuh}\\
 \hline
 \hline
\GOSAM{} + {\sc Herwig++/Matchbox}\,\cite{Bellm:2013lba,Platzer:2011bc}\\
\hline
$pp\to Z/\gamma^*$+1 jet (+shower)\,\cite{Butterworth:2014efa}\\
\hline
\end{tabular}
\end{center}
\end{table}

\subsection{Examples Beyond the Standard Model}
\label{sec:BSMexample}

Examples for the usage of \GOSAM{} with a model file different
from the Standard Model are the calculation of the SUSY QCD corrections to 
neutralino pair production in associatin with an extra jet, including full 
off-shell effects~\cite{Cullen:2012eh}, or 
the QCD corrections to graviton plus jet 
production in models with large extra dimensions~\cite{Greiner:2013gca}.

The corresponding model files in {\tt UFO}\,\cite{Degrande:2011ua} format 
were generated using {\tt FeynRules}\,\cite{Christensen:2008py}. 
To import new model
files within the \GOSAM{} setup, the user only needs to specify the path to the
model file in the {\it process card}.  
For more details about the setup and the phenomenological results we refer to \cite{Cullen:2014yla,Cullen:2012eh,Greiner:2013gca}.
